\documentstyle[12pt]{article}
\begin{document}
\thispagestyle{empty}
\mbox{ }\hfill{\normalsize UCT-TP 191/93}\\
\mbox{ }\hfill{\normalsize 27 January 1993}\\
\begin{center}
{\Large \bf Indications for a\\[.5cm]
Detonating Quark-Gluon Plasma}\\[1.cm]
{\bf N. Bili\'c$^{1,2}$,
J. Cleymans$^1$,
E. Suhonen$^{1,3}$,
and D.W. von Oertzen$^4$}\\[2.0em]
$^1$Department of Physics,
University of Cape Town, Rondebosch 7700, South Africa  \\
$^2$Rudjer Bo\v{s}kovi\'{c} Institute,
P.O. Box 1016, Zagreb, Croatia                          \\
$^3$Department of Theoretical Physics,
University of Oulu, SF-90550 Oulu, Finland  \\
$^4$Physics Department,
University of Namibia, Windhoek, Namibia
\end{center}
\vskip 1cm
\begin{abstract}
We propose a mechanism which naturally contains the relation
$\mu_{B} = 3\mu_{S}$ of the hadronic gas produced in heavy-ion
collisions at CERN. Our starting assumption is the existence of a
sharp front separating the quark-gluon plasma phase from the hadronic
phase. Energy-momentum conservation across the front leads to the
following consequences for an adiabatic process
\begin{description}
\item[{\rm a)}]  The baryon chemical potential, $\mu_{B}$, is
               approximately continuous across the front.
\item[{\rm b)}] The temperature in the hadronic gas is higher than the
               phase transition temperature due to superheating.
\item[{\rm c)}] In the region covered by the experiments the velocity
               of the hadronic gas approximately equals the speed of
               sound in the hadronic gas.
\end{description}
\end{abstract}
\newpage

Several groups
\cite{ref1}-\cite{refpretoria3}
 have investigated the  results
obtained by the CERN relativistic ion experiments
on the production of strange particles
\cite{refNA35}-\cite{refNA44}.
These results are very sensitive to the
particular kinematic window in which particle yields are being measured.
The results from the WA85 group \cite{refWA85}
have attracted particular attention because
they measure several of the strange baryons in a very narrow kinematic
region, central in rapidity and at large transverse momenta. Within the
framework of hadronic gas models these experimental results lead to the
conclusion
 that the hadrons were produced from a system having a
temperature $T$ of approximately 200 MeV, a baryon chemical
potential $\mu_B$ of about
300 MeV and a strangeness chemical
potential $\mu_S$ of 100 MeV
\cite{ref1,refrafelski1,refpretoria3}.
The ratio of strange to baryon chemical potential thus obtained is
particularly intriguing as the relation
$$
\mu_B=3\mu_S
\eqno(1)
$$
naturally holds in a quark-gluon plasma \cite{ref1}
guaranteeing strangeness neutrality between strange and
anti-strange quarks. In a hadronic gas
there is however no particular reason for such a relation to hold.
Early proposals to explain it have difficulties with the
entropy generated in the transition and violate the second law of
thermodynamics~\cite{ref2}.

In the present paper we propose a model which is obtained
naturally if one assumes the existence of a sharp front separating the
hadronic and quark-gluon plasma phases.
Ahead and behind the front each phase is behaving as an (approximately)
free gas of particles, except for the fact that we take into account the
hard core of the baryons and the presence of resonances.
All the dynamics
related to the  transition takes place inside this sharp front.
In going from one
phase to the other we take into account energy-momentum conservation. This
provides for a certain relationship between the energy-density in the
hadronic phase and in the quark-gluon phase.
 In order for the transition to be
possible, a substantial amount of supercooling has to happen in the
quark-gluon phase while a super-heated hadronic gas emerges on the other
side. This way there is no problem with the entropy since the supercooled
quark-gluon gas will have a low entropy content.
One is then led to the picture which is
shown in figure~1.
Such a transition is well
known in hydrodynamics and is usually referred to as a detonation
transition (see e.g. Ref. \cite{refLandau}). The idea of a sharp front
was first put forward in this context by Van Hove \cite{refVanhove}, it was
considered in more detail later on by several groups
\cite{refallbigshots}-\cite{refseibert}.
It was also considered recently in the context of particle production in
Ref. \cite{refrussians}. It has however not been used in any systematic
way to make contact with the recent experimental results obtained with the
CERN relativistic ion experiments. This paper is a first step towards such
an analysis.

The attractive features of our model are as follows :
\begin{enumerate}
\item The baryon chemical potential $\mu_B$ is
continuous across the front to a very good approximation.
\item The relation $\mu_B=3\mu_S$ is exactly
preserved during the transition.
\item The temperature in the hadronic gas is higher than the phase
transition temperature due to the resulting superheating. This would
explain why numerical simulations
\cite{refkogut}-\cite{refneven}
of lattice QCD  obtain estimates for the
critical temperature which are lower than the temperature observed in the
transverse momenta spectra measured in relativistic ion
collisions.
\item The hadronic gas emerges approximately with the
velocity of sound.
\end{enumerate}

Imposing the condition $\mu_B=3\mu_S$ in the hadronic phase
constrains the allowed values of the temperature.
Normally $\mu_S$ is a free parameter whose value is fixed by the condition
of strangeness neutrality in the hadronic gas. Fixing it to the
above relation implies imposing a restriction on the temperature. In
fact,  the hadronic gas cannot overheat beyond a
certain limit (about 205 MeV in our case).
The resulting curve in
the $T-\mu_B$ plane is shown in figure 2. Note also that for  $\mu_B=0$
no constraint exists since $\mu_S$ is then also zero and the relation
is trivially satisfied.

As stated before we assume the existence of a sharp front separating the
two phases.
The standard analysis considers
a one-dimensional situation in the frame of reference where the front is
at rest (see e.g. Ref. \cite{refLandau}).
Energy and momentum conservation across the front
then lead to the two following equations :
$$
(\epsilon_h + P_h)\gamma_h^2 v_h + P_h =(\epsilon_q + P_q)\gamma_q^2 v_q +
P_q ,
\eqno(2)
$$
$$
(\epsilon_h+P_h)v_h^2\gamma_h^2=(\epsilon_q+P_q)v_q^2\gamma_q^2 ,
\eqno(3)
$$
both sides of the front are at  different temperatures and
chemical potentials. The thermodynamic quantities (pressure, energy
density) refer to the rest frame quantities.
The index $q$ refers to the quark-gluon plasma phase while the index $h$
refers to the hadronic phase, $v$ refers to the velocity
of the gas in each phase with
respect to the front, $\gamma$ is the standard relativistic factor
($1/\sqrt{1-v^2})$.
In addition to the above two equations, baryon number is also conserved
across the front which leads to
$$
n_hv_h\gamma_h=n_qv_q\gamma_q.
\eqno(4)
$$
where $n_h$ and $n_q$ refer to the baryon densities
in the hadronic and in the quark-gluon plasma phases respectively.
The above equations can be solved for the velocities, leading to
$$
v_q^2={(P_h-P_q)(\epsilon_h+P_q)\over
(\epsilon_h-\epsilon_q)(\epsilon_q+P_h)},
\eqno(5)
$$
$$
v_h^2={(P_h-P_q)(\epsilon_q+P_h)\over
(\epsilon_h-\epsilon_q)(\epsilon_h+P_q)} .
\eqno(6)
$$
The change in density can then be determined from
$$
\left({n_q\over n_h} \right)^2=
{(\epsilon_q+P_q)(\epsilon_q+P_h)\over
(\epsilon_h+P_h)(\epsilon_h+P_q)}.
\eqno(7)
$$
Another important constraint
is given by the requirement of non-decreasing entropy
which implies in our notation
$$
s_hv_h\gamma_h\geq s_qv_q\gamma_q .
\eqno(8)
$$
Combining (4) and (8) leads to
$$
{s_h\over s_q} \geq {n_h\over n_q} ,
\eqno(9)
$$
The equality sign is of particular interest since it corresponds to an
adiabatic transition.

To exploit these relations fully we have to specify the equation of state
(EOS) used in each phase. On the quark gluon plasma side we use the bag EOS
given by
$$
\epsilon_q=3P_q+4B ,
\eqno(10)
$$
where $B$ is the bag  constant. We have chosen $B=0.2$ GeV/fm$^3$.

On the hadronic side we incorporate the hard core radius of hadrons by
choosing \cite{refhelmut}
$$
\epsilon_h={\epsilon_h^0\over 1+V_0n_h^0} ,
\eqno(11)
$$
where the subscript zero refers to point-like quantities, $V_0$ refers to
a typical proper volume of a hadron which we take to have a radius given
by 0.8 $fm$.
The hadronic side in our approach is a composition of non-interacting
quantum gases. The use of quantum statistics is essential
since in a wide range
of $\mu_B$ and $T$ the Boltzmann approximation is not satisfied.
We include  all
well established hadronic resonances listed in
the latest edition of the
Review of Particle Properties \cite{refParticles}. A
correction similar to equation (11) is performed to all thermodynamic
quantities.

We will pay particular attention to adiabatic transitions corresponding to
$$
{s_q\over n_q}={s_h\over n_h} ,
\eqno(12)
$$
i.e. where the entropy per baryon remains constant across the transition.
To determine the temperature in the quark phase we proceed as follows.
Introducing the ratio $z$ defined by
$$
z=\mu_q/(\pi T_q) ,
\eqno(13)
$$
one can use
the equation of state described above for the quark-gluon phase to
obtain
$$
{s_q\over n_q} = {3\pi\over z}\left( 1-{11\over 180 (1+z^2)}\right) .
\eqno(14)
$$

We have chosen the number of flavours equal to 2.5 to simulate the effect
of the strange quark mass while keeping the expressions for massless
quarks \cite{refkajantie}.
This determines $z$ for an adiabatic transition in terms of the hadronic
temperature and baryon chemical potential. From the ratio of baryon
densities given by equation (6) one can then determine the temperature
$T_q$ in the quark-gluon phase.

In figure 3 we show the energy densities in the quark
and in the hadronic phases as  functions of the temperature $T_h$.
One sees that the energy density
is always lower in the
quark sector than in the hadronic sector for the range of temperatures
considered.

In figure 4 we show the corresponding temperature in the quark phase. As
one can see, it is always very low, indicating a substantial amount of
supercooling in the quark phase.

Figure 5 is probably the most interesting result of this analysis. In this
case we show the baryon chemical potential in the hadronic phase and the
corresponding one on the quark side
(denoted by $3\mu_Q$) varying the bag constant in the range 0.16-0.24
GeV/fm$^3$.
The two chemical potentials are nearly equal. This
means that the baryon chemical potential is conserved through the front
even though the temperatures are very different.   So we have the
interesting situation where all the thermodynamic quantities change quite
drastically during the transition but the chemical potentials remain
almost unaffected.

The velocities near the front have a special hydrodynamic importance.
As is well known, small perturbations in a medium  propagate with the
velocity of sound in that medium. It is therefore of special interest to
compare $v_q$ and $v_h$ with the velocity of sound in each phase.
The latter is defined as \cite{refLandau}
$$
c_s^2=\left.{\partial P\over\partial\epsilon}\right|_{s/n} .
\eqno(15)
$$
In the quark-gluon phase, eq. (10) yields $c_{s,q}^2=1/3$.
 To calculate the speed of sound in the hadronic gas
we write eq. (15) as
$$
c_{s,h}^2={\partial P_h/\partial T +\partial P_h/\partial\mu_B\cdot d\mu_B/dT
      \over
 \partial\epsilon_h/\partial T +\partial\epsilon_h/\partial\mu_B\cdot
d\mu_B/dT} ,
\eqno(16)
$$
where the  derivative
$d\mu_B/dT$ is taken such that the ratio $s_h/n_h$ remains constant. Thus
$$
{d\mu_B\over dT}=-{n_h\partial s_h/\partial T -s_h\partial n_h/\partial T
                 \over
                 n_h\partial s_h/\partial\mu_B -s_h\partial
n_h/\partial\mu_B} .
\eqno(17)
$$

The velocities thus calculated are shown in figure 6
together with $v_q$ and $v_h$. It can be seen that
the quark-gluon plasma velocity is very high, quarks stream in at  high
speed but with a low density. On the other side of the front the
hadrons are slower  however with a  larger density.

It is interesting to note that around the relevant values of $T$ and
$\mu_B$ namely 200 and 300 MeV respectively, the hadrons emerge with sound
velocity, while the quarks have a velocity which is considerably larger
than the speed of sound in a quark-gluon plasma.

One also observes in this model a large increase in the entropy density as
one makes the transition from a supercooled quark-gluon phase to a
superheated hadronic phase.

In summary, we have considered a model based on the existence of a sharp
front separating the hadronic from the quark-gluon phase.
We arrive at the conclusion
that in this situation the transition can only take place if a substantial
amount of super-cooling takes place in the quark-gluon phase and that the
hadronic phase appears in a super-heated state. This would explain why
such high temperature have been observed experimentally while  the
data from lattice QCD indicate that the phase transition
temperature is considerably lower than 200 MeV.
It is interesting to note that the
chemical potentials remain approximately the same in such a transition
 while the temperature
changes drastically in the transition.
The temperature on the hadronic
side is about 100 MeV higher  than on the quark-gluon side. In the frame
where the sharp front is at rest the velocity of the quark phase is above
the corresponding sound velocity while in the hadronic phase it is very
close to the hadronic speed of sound.
\vskip 0.5 cm
\noindent
{\bf Acknowledgement} We acknowledge
continuing stimulating discussions in this field
with H. Satz and K. Redlich. Two of us (E.S. and D.W.v.O.) thank the Physics
Department of the University of Cape Town for its hospitality.

\newpage
\noindent {\bf Figure Captions:}
\begin{enumerate}
\item Schematic diagram of the energy-density as a function of
temperature. A detonation transition from a supercooled quark-gluon
plasma (long dashed line) to a superheated hadron gas (short dashed line) is
indicated.
\item The hadronic temperature $T_h$ as a function of the baryon
chemical potential $\mu_B$.
\item The energy-density of the hadronic gas (full line)
and of the quark-gluon plasma (dashed line) as a function
of the hadronic temperature.
\item The quark temperature $T_q$ versus the hadronic temperature
$T_h$.
\item The baryon chemical potential in the quark-gluon plasma
for $B=0.24$ GeV/fm$^3$ (dash-dotted line), for $B=0.2$ GeV/fm$^3$
(fullline), for $B=0.16$ GeV/fm$^3$ (long dashes)
and in the hadronic gas (short dashes).
\item The velocity squared of the quark matter(long dashed line) and of
the hadronic matter (short dashed line) relative to the transition front.
Also shown are the hadronic
sound velocity squared with (full line) and without
(dot-dashed line) excluded volume corrections.
\end{enumerate}
\end{document}